\documentclass[a4paper]{aa}
\usepackage{graphics,txfonts}

\def\teff{$T_{\rm eff}$}
\def\lgg{$\log g$}

\def\vs{$v_{\rm e}\sin i$}

\newcommand{\bs}{$\langle B_{\rm s}\rangle$}
\newcommand{\kms}{km\,s$^{-1}$}

\newcommand{\ei}{E$_{\rm i}$}

\newcommand{\roap}{$^*$}
\newcommand{\noap}{\phantom{$^*$}}

\newcommand{\txtbf}[1]{#1}

\newcommand{\figps}[1]{\resizebox{\hsize}{!}{\rotatebox{0}{\includegraphics{#1}}}}

\newcommand{\fifps}[2]{\centering\resizebox{#1}{!}{\includegraphics{#2}}}

\newcommand{\figgps}[2]{\resizebox{#1}{!}{\rotatebox{0}{\includegraphics{#2}}}}

\begin{document}

\title{Isotopic anomaly and stratification of Ca in magnetic Ap stars%
\thanks{Based on observations collected at the European Southern Observatory,
Paranal, Chile (ESO program No. 68.D-0254)}}

\author{T. Ryabchikova \inst{1,2} \and 
        O. Kochukhov   \inst{3} \and
        S. Bagnulo     \inst{4}}
 
\offprints{T. Ryabchikova, \\ \email{ryabchik@inasan.ru}}

\institute{Department of Astronomy, University of Vienna, T\"urkenschanzstra{\ss}e 17, 1180 Vienna, Austria
      \and Institute of Astronomy, Russian Academy of Sciences, Pyatnitskaya 48, 109017 Moscow, Russia
      \and Department of Physics and Astronomy, Uppsala University, SE-751 20, Uppsala, Sweden
      \and Armagh Observatory, College Hill, Armagh BT61 9DG, Northern Ireland}

\date{Received / Accepted }

\abstract{}
{
 We have carried out an accurate investigation of the Ca isotopic composition
 and stratification in the atmospheres of 23 magnetic chemically
 peculiar (Ap) stars of different temperature and magnetic field strength.
}
{
 With the UVES spectrograph at the 8\,m ESO VLT, we have
 obtained high-resolution spectra of Ap stars in the wavelength range
 3000--10000\,\AA.  Using a detailed spectrum synthesis
 calculations, we have reproduced a variety of Ca lines in the optical and
 ultraviolet spectral regions, inferring the overall vertical
 distribution of Ca abundance, then we have deduced the relative
 isotopic composition and its dependence on height using the profile of
 the the IR-triplet \ion{Ca}{ii} line at $\lambda$~8498\,\AA.
}
{ 
 In 22 out of 23 studied stars, we found that Ca is strongly stratified, being 
 usually overabundant by 1.0--1.5\,dex below $\log\tau_{5000}\approx -1$,
 and strongly depleted above $\log\tau_{5000}=-1.5$.  The IR-triplet
 \ion{Ca}{ii} line at $\lambda$~8498\,\AA\ reveals a significant
 contribution of the heavy isotopes $^{46}$Ca and $^{48}$Ca, which
 represent less than 1\,\% of the terrestrial Ca isotopic mixture. We
 confirm our previous finding that the presence of heavy Ca isotopes
 is generally anticorrelated with the magnetic field
 strength. Moreover, we discover that in Ap stars with relatively
 small surface magnetic fields ($\le$\,4--5 kG), the light isotope
 $^{40}$Ca is concentrated close to the photosphere, while the heavy
 isotopes are dominant in the outer atmospheric layers. This vertical
 isotopic separation, observed for the first time for any metal in a
 stellar atmosphere, disappears in stars with magnetic field strength
 above 6--7\,kG.
}
{
 We suggest that the overall Ca stratification and depth-dependent
 isotopic anomaly observed in Ap stars may be attributed to a combined
 action of the radiatively-driven diffusion and light-induced drift.
}

\maketitle

\section{Introduction}
\label{intro}
 
After the pioneering work by Michaud (\cite{Michaud70}), atomic diffusion in stellar envelopes and
atmospheres has been recognized as the main process responsible for the atmospheric abundance
anomalies in the peculiar stars of the upper main sequence. Due to their unique characteristics,
such as an extremely slow rotation, strong, global magnetic fields, and the absence of convective 
mixing, magnetic chemically peculiar (Ap and Bp) stars  exhibit the most clear manifestation of
diffusion effects and thus represent privileged laboratories for investigation of chemical
transport processes and magnetohydrodynamics. 

Detailed diffusion calculations performed for a set of chemical elements in the atmospheres of
magnetic peculiar stars predict separation of chemical elements over the stellar surface and with
height in stellar atmosphere (abundance stratification). These theoretical predictions can be
directly tested through the comparison with empirical maps of chemical elements inferred from
observations. For a small number of elements, including Ca, an effect of the vertically stratified
element distribution on the spectral line profiles was demonstrated in early studies (Borsenberger
at al. \cite{BPM81}). However, the absence of high-resolution, high signal-to-noise spectroscopic
observations did not allow a robust comparison between observations and theoretical diffusion
modelling. This step was carried out by Babel (\cite{Babel92}), who calculated the Ca abundance
distribution in the atmosphere of magnetic Ap star 53~Cam and showed that the unusual shape of
\ion{Ca}{ii} K line -- a combination of the wide wings and extremely narrow core 
(Babel \cite{Babel94}; Cowley et al. \cite{CHK06}) -- is a
result of a step-like Ca distribution with abundance decrease at $\log\tau_{5000}\approx-1$.
Following Babel, the step-function approximation for the abundance distributions was employed in
many stratification studies based on the observed profiles of spectral lines
(\txtbf{Savanov et al. \cite{SKV01}};
Bagnulo et al. \cite{bagetal01}; Wade et al. \cite{WLRK03}; 
Ryabchikova et al. \cite{RPK02,RLK05,RRKB06}; \txtbf{Glagolevskii et al. \cite{GRC05};
Cowley et al. \cite{CHC07}}).

Ca was found to be stratified in nearly the same way as in 53 Cam (enhanced concentration of Ca
below $\log\tau_{5000}\approx-1$ and its depletion above this level) in all stars for which
stratification analysis have been performed: $\beta$~CrB (Wade et al. \cite{WLRK03}), $\gamma$~Equ
(Ryabchikova et al. \cite{RPK02}), HD~204411 (Ryabchikova et al. \cite{RLK05}), HD~133792 
(Kochukhov et al. \cite{KTRM06}) and HD~144897 (Ryabchikova et al. \cite{RRKB06}).  In addition to
remarkable diffusion signatures, recently another Ca anomaly was detected, first in the spectra of
the Hg-Mn stars by Castelli \& Hubrig (\cite{CastH04}) and then in Ap stars by Cowley \& Hubrig
(\cite{CH05}). These authors found a displacement of the lines of \ion{Ca}{ii} IR triplet due to 
significant contribution of the heavy Ca isotopes. 
Ryabchikova, Kochukhov \& Bagnulo (see review paper by Ryabchikova (\cite{MONS05})) were 
the first to apply spectrum synthesis calculations to investigate effects of Ca isotopes on the 
calcium line profile shape in magnetic CP stars.
In particular, we have demonstrated the general anticorrelation
between the presence of heavy Ca and magnetic field strength: in Ap stars the
contribution of heavy Ca isotopes decreases with the increase of the field modulus, and disappears
when the field strength exceeds $\sim$\,3\,kG.
Cowley et al. (\cite{CHC07}) have come to a similar conclusion, but with reservations.

In the present paper we summarise our detailed analysis of the vertical stratification of Ca
abundance in the atmospheres of magnetic Ap stars of different temperatures and
magnetic field strengths. We combine this vertical Ca mapping with the analysis
of the Ca isotopic anomaly and its dependence on height, based on the spectrum synthesis modelling
of the IR triplet \ion{Ca}{ii} line at $\lambda$~8498\,\AA. Our study is the first to present a
homogeneous and systematic determination of the vertical stratification of a given element in
the atmospheres of a large number of stars, enabling us to study the signatures of 
atmospheric atomic diffusion in the presence of strong magnetic field as a function of stellar 
parameters and magnetic field intensity.

Our paper is organized as follows. In Sect.~\ref{obs} we describe spectroscopic
observations and  the data reduction. Determination of the stellar atmospheric
parameters is detailed in Sect.~\ref{parameters}. Sects.~\ref{synthesis}--\ref{isot}
present our methodological approach to magnetic spectrum synthesis, determination of
Ca stratification and the study of Ca isotopic composition. Our findings are
summarised in Sect.~\ref{results} and are discussed in Sect.~\ref{discus}.

\section{Observations and data reduction} 
\label{obs}

Twenty-three slowly rotating Ap stars were chosen for the Ca stratification analysis. 
The list of program stars is given in Table~\ref{tbl1}. In addition, 
three stars, Procyon (=HD~61421, spectral type F5), HD~27411
(=HIP~20106, spectral type A3m), and HD~73666 (=40 Cnc, spectral type A1\,V) 
were used as standards for the Ca isotopic study.
Of the 23 magnetic Ap stars included in our study, 12 stars show 
high-overtone acoustic {\it p-}mode pulsations and thus belong to the group of rapidly 
oscillating Ap (roAp) stars.

We note that metallic line stars show signatures of strong turbulence and convection in  their
atmospheres (Landstreet \cite{L98}; Kochukhov et al. \cite{KFP06}). Theoretical diffusion
models for Am stars (e.g. Michaud et al. \cite{MRR05}) achieve reasonable success in reproducing the
surface abundance patterns assuming full mixing of the outer envelope. Thus, despite non-solar
abundances, strong mixing inhibits development of the vertical chemical gradients in the
line-forming region and for this reason Am star HD~27411 can be considered as a comparison object
in the context of the present investigation.

For the majority of our targets high-resolution, high signal-to-noise-ratio spectra were obtained
with the UVES instrument (Dekker et al. \cite{DOK00}) at the ESO VLT in the context  of program
68.D-0254. The observations were carried out using both available dichroic modes (standard
settings 346+580\,nm and 437+860\,nm). In both the blue arm and the red arm the slit width was set
to 0.5$^{\prime\prime}$, for a spectral resolution of about 80\,000. The slit was oriented along
the parallactic angle, in order to minimize losses due to atmospheric dispersion.  Almost the full
wavelength interval from 3030 to 10400\,\AA\ was observed except for a few gaps, the largest of
which was at 5760--5835\,\AA\ and 8550--8650\,\AA. In addition, there are several small gaps, about
1\,nm each, due to the lack of overlapping between the \'{e}chelle orders in the 860U setting. 

Spectra of HD~24712 and HD~61421 (Procyon) were acquired with the UVES at VLT as part of the
UVESpop project\footnote{{\tt http://www.eso.org/uvespop/}} (Bagnulo et al. \cite{uvespop}).
The UVES observations of the magnetic Ap star
HD\,66318 employed in our paper were described by Bagnulo et al. (\cite{bagn03}). All three stars
were observed using the same instrumental settings as the Ap targets from our main data set.
Spectrum of HD~73666 was obtained with the ESPaDOnS spectropolarimeter at the Canada-France-Hawaii
Telescope and analysed for abundances by Fossati et al. (\cite{fossati}). A reduced spectrum was
kindly provided to us by the authors.  

The UVES spectra have been reduced with the automatic pipeline described in Ballester et al.
(\cite{BMB00}). For all settings, science frames are bias-subtracted and divided by the extracted
flat-field, except for the 860~nm setting, where the 2-D (pixel-to-pixel) flat-fielding is used,
in order to better correct for the fringing. Because of the high flux of the spectra, we used the
UVES pipeline \textit{average extraction} method.

Due to the gaps in spectral coverage, only one line of the \ion{Ca}{ii} IR triplet,
$\lambda$~8498\,\AA, could be observed in all stars and is accessible for modelling. This line is
the weakest among the three IR triplet transitions, which facilitates quantitative
analysis, especially in cooler stars where the triplet \ion{Ca}{ii} lines become particularly 
strong. 

All three lines of the \ion{Ca}{ii} IR triplet overlap with the hydrogen Paschen lines, which
become very prominent in the spectra of hotter Ap stars (see Fig.~\ref{hbop}).  The 8498\,\AA\ line
is  located further away from the centre of the nearest Paschen line (P16) than the other two
lines of the IR triplet. The hydrogen line blending for this \ion{Ca}{ii} transition is also
weaker than for the other two \ion{Ca}{ii} lines. 
Moreover, being the weakest among three IR triplet lines the 8498\,\AA\ line is less dependent
on the accuracy of  Stark and Van der Waals broadening parameters.
These circumstances suggest the \ion{Ca}{ii} 8498\,\AA\ line is most suitable for detailed modelling.

The studied \ion{Ca}{ii} line lies rather close to the gap in the wavelength coverage of the UVES
setting employed. An additional difficulty arises due to the broad overlapping absorption produced
by the Paschen lines. These properties of the data complicate continuum normalization in the
region around \ion{Ca}{ii} 8498\,\AA. In fact, with the available data, we are generally unable to
rectify the spectra in the usual manner. Instead, for each star observations were adjusted to
match calculated Paschen line spectrum in a small region around the P16 line. This procedure was
carried out in two steps. First, a high-order spline function was employed to trace the wings of
the hydrogen line, and observations were divided by this fit. Secondly, observed spectra were
multiplied by the theoretical hydrogen line spectrum calculated using individual stellar
atmospheric parameters and hydrogen line opacity described in Sect.~\ref{synthesis}.

For most program stars spectral coverage of our UVES spectra precluded analysis of the two
stronger lines of the IR triplet, \ion{Ca}{ii} 8542 and 8662~\AA. However, for one of the targets,
HD\,24712, we have investigated these lines using complimentary observations obtained with the
ESPaDOnS instrument at CFHT. The ESPaDOnS spectrum HD\,24712 is the average of 81 time-resolved
observations of this star analysed by Kochukhov \& Wade (\cite{KW07}). We refer to this paper for
the details of reduction of these data. Unlike the UVES spectrum of HD\,24712, the ESPaDOnS data were 
collected close to the phase of magnetic maximum (JD 2453744.71--2453744.83).

\begin{table*}[!th]
\caption{Fundamental parameters of the program stars. 
The columns give the target HD number, range of Julian dates of its observation with UVES,
effective temperature, surface gravity, projected rotational velocity,
estimate of the mean magnetic field modulus and the ratio of radial to azimuthal 
field strength adopted in spectrum synthesis. Rapidly oscillating Ap (roAp) stars are identified by 
asterisk. \label{tbl1}}
\begin{center}
\begin{tabular}{rclccccl}
\hline
\hline
HD~~~~  & JD$-$2450000 &~~\teff     & \lgg & \vs~~     & \bs & $B_{\rm r}/B_{\rm a}$ & ~~Reference  \\
number~ &              &~~(K)	 &	& (\kms)    & (kG)&		\\
\hline
\multicolumn{7}{c}{\it Magnetic chemically peculiar stars} \\
   965\noap &2190.632--2190.646&~~7500& 4.00 & ~3.0   & 4.4 & 3.2/3.2   &RKP   \\
 24712\roap &1982.545--1982.561&~~7250& 4.30 & ~5.6   & 2.3 & 2.3/0.0   &RLG97 \\  
 29578\noap &2213.805--2213.819&~~7800& 4.20 & ~2.5   & 5.6 & 4.5/3.4   &RNW04 \\  
 47103\noap &2286.690--2286.719&~~8180& 3.50 & ~0.0   &16.3 & 0.0/16.3  &RKP   \\                         
 66318\noap &2413.467--2413.477&~~9200& 4.25 & ~0.0   &15.5 & 10.2/11.7 &BLL03 \\    
 75445\noap &2236.837--2236.846&~~7650& 4.00 & ~3.0   & 3.0 & 2.8/1.0   &RNW04 \\                         
101065\roap &2280.849--2280.863&~~6600& 4.20 & ~3.5   & 2.3 & 2.3/0.0   &CRK00 \\
111133\noap &2294.861--2294.866&~~9930& 3.65 & ~5.0   & 4.0 & 4.0/0.0   &RKP   \\
116114\roap &2296.864--2296.871&~~8000& 4.10 & ~2.5   & 6.2 & 4.6/3.9   &RNW04 \\  
118022\noap &2298.861--2298.865&~~9500& 4.00 & 10.0   & 3.0 & 3.0/0.0   &RKP   \\                        
122970\roap &2295.862--2295.875&~~6930& 4.10 & ~5.5   & 2.5 & 2.5/0.0   &RSH00 \\  
128898\roap &2320.870--2320.873&~~7900& 4.20 & 12.5   & 1.5 & 1.5/0.0   &KRW96 \\        
133792\noap &2331.791--2331.797&~~9400& 3.70 & ~0.0   & 1.1 & 1.1/0.0   &KTR06 \\  
134214\roap &2331.816--2331.826&~~7315& 4.45 & ~2.0   & 3.1 & 2.5/1.7   &RKP   \\                        
137909\roap &2331.892--2331.895&~~8000& 4.30 & ~2.5   & 5.4 & 5.0/2.0   &RNW04 \\  
137949\roap &2331.803--2331.810&~~7550& 4.30 & ~1.0   & 5.0 & 2.2/4.5   &RNW04 \\  
144897\noap &2331.834--2331.851& 11250& 3.70 & ~3.0   & 8.8 & 6.3/6.2   &RRK06 \\ 
166473\roap &2189.508--2189.517&~~7700& 4.20 & ~0.0   & 8.6 & 5.0/7.0   &GRW00 \\   
170973\noap &2190.511--2190.518& 10750& 3.50 & ~8.0   & 0.0 &           &K03   \\             
176232\roap &2190.522--2190.528&~~7650& 4.00 & ~2.0   & 1.5 & 1.5/0.0   &RSH00 \\               
188041\noap &2190.599--2190.605&~~8800& 4.00 & ~0.0   & 3.6 & 3.4/1.0   &RLK04 \\ 
203932\roap &2189.540--2189.557&~~7550& 4.34 & ~5.3   &$\le$1&          &GKW97 \\
217522\roap &2189.567--2189.577&~~6750& 4.30 & ~2.5   &$\le$1.5&1.5/0.0 &G98   \\
\multicolumn{7}{c}{\it Comparison stars} \\
 27411\noap &2501.937--2501.946&~~7650& 4.00 & 18.5   & 0.0 & &RKP  \\
 61421\noap &2555.906--2555.909&~~6510& 3.96 & ~3.5   & 0.0 & &AAG02\\
 73666\noap &3745.063--3745.082&~~9382& 3.78 & 10.0   & 0.0 & &FBM07  \\
\hline
\end{tabular}
\end{center}
References. RKP -- this paper; RLG97 -- Ryabchikova et al. (\cite{RLG97}); 
RNW04 -- Ryabchikova et al. (\cite{RNW04}); BLL03 -- Bagnulo et al. (\cite{bagn03});
CRK00 -- Cowley et al. (\cite{CRK00}); RSH00 -- Ryabchikova et al. (\cite{RSH00});
KRW96 -- Kupka et al. (\cite{KRW96}); KTR06 -- Kochukhov et al. (\cite{KTRM06});
RRK06 -- Ryabchikova et al. (\cite{RRKB06}); GRW00 -- Gelbmann et al. (\cite{GRW00});
K03 -- Kato (\cite{Kato03}); RLK04 -- Ryabchikova et al. (\cite{RLKB04});
GKW97 -- Gelbmann et al (\cite{GKWM97}); G98 -- Gelbmann (\cite{G98});
AAG02 -- Allende Prieto et al. (\cite{Procyon}); FBM07 -- Fossati et al. (\cite{fossati}).
\end{table*}

\section{Fundamental stellar parameters}
\label{parameters}

Fundamental parameters of the program stars are given in Table~\ref{tbl1}. For most stars
effective temperatures \teff\ and surface gravities \lgg\ were taken from the recent studies
(references are provided in the last column of Table~\ref{tbl1}).  For HD\,965, HD\,47103,
HD\,118022 and HD\,134214 atmospheric parameters were derived using the Str\"omgren photometric
indices extracted from the catalogue of  Hauck \& Mermilliod (\cite{HM98}). We used calibrations
by Moon \& Dworetsky (\cite{MD85}) and by Napiwotzki et al. (\cite{N93}) as implemented in the
{\tt TEMPLOGG} code (Rogers \cite{R95}). For HD\,111133 the effective temperature was taken from
Kochukhov  \& Bagnulo (\cite{KB06}), while other parameters were derived in the present study.  
For HD\,75445, HD\,176232 and
HD\,203932 effective temperatures were further refined by fitting the H$\alpha$ profile.
This resulted in a small correction relative to the published \teff\ values.

The mean surface magnetic fields \bs\ were derived using spectral lines which exhibit resolved and
partially resolved Zeeman splitting patterns.  In all stars rotational velocities were estimated
by modelling profiles of the magnetically insensitive \ion{Fe}{i} lines at $\lambda\lambda$~5434.5
and 5576.1\,\AA.  Model atmospheres for all program stars were calculated with the {\tt ATLAS9}
code  (Kurucz \cite{K93}), employing opacity distribution functions (ODFs) with an enhanced
metallicity. For stars with especially strong magnetic fields we used ODFs with a non-zero 
\txtbf{pseudo-microturbulence} velocity in order to simulate the modification of the line opacity due to 
magnetic intensification of spectral lines. All spectrum synthesis for
magnetic stars employed zero microturbulence. For comparison stars we have adopted the following
values: $\xi_{\rm t}=1.8$~\kms\ for Procyon (Allende Prieto et al. \cite{Procyon}),
1.9~\kms\ for HD\,73666 (Fossati et al. \cite{fossati}) and 2.5~\kms\ for HD\,24711 (determined
in this paper).

\section{Spectrum synthesis calculations}
\label{synthesis}

Analysis of the Ca stratification and isotopic composition was based on detailed magnetic spectrum
synthesis calculations with the {\tt SYNTHMAG} code (Kochukhov \cite{synthmag06}). This program allows
to solve numerically polarized radiative transfer equation and calculate theoretical stellar
spectra in four Stokes parameters for the prescribed model atmosphere and vertical distribution of
any number of chemical elements. The {\tt SYNTHMAG} calculations performed in this paper are based
on a simplified model of the stellar magnetic field topology, characterized by a single value of the
field modulus and a homogeneous field distribution over the surface of the star. 
When rotation broadening is significant or the field is too weak, we assume purely radial 
field orientation. In other cases the effective field orientation is inferred from the profiles of
magnetically sensitive lines and is parameterized with the ratio of the radial to azimuthal field 
components (see Table~\ref{tbl1}). Although
undoubtedly rather simplified, this magnetic model is quite successful in explaining the
shapes of resolved Zeeman split lines in the Stokes $I$ spectra of many Ap stars
\txtbf{(Kochukhov et al. \cite{KLR02}; Nielsen \& Wahlgren \cite{NW02};
Leone et al. \cite{LVS03}; Ryabchikova et al. \cite{RRKB06})}
and, in some
cases, performs even better than low-order multipolar field geometries \txtbf{(Kochukhov
\cite{K07b})}. 

\begin{figure}[!t]
\figps{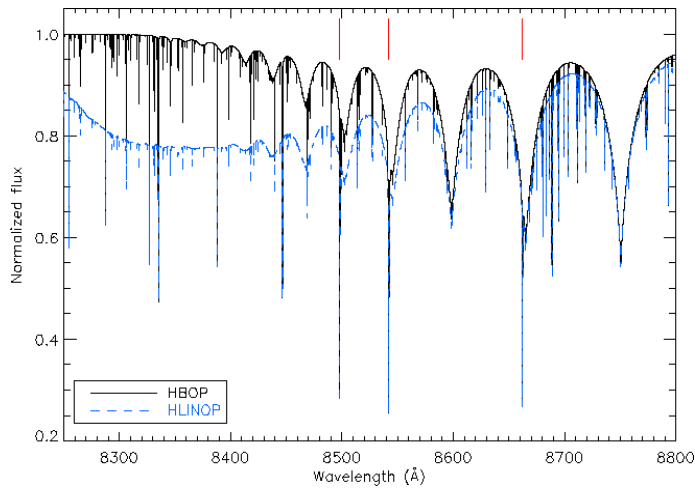}
\caption{Synthetic spectra calculated with different treatment of the overlapping hydrogen
line opacity. Usual calculation ({\tt HLINOP} routine, dashed line) is compared with
spectrum synthesis based on the occupation probability formalism
({\tt HBOP} routine, solid line). In both cases {\tt ATLAS9} model atmosphere with \teff\,=\,8500~K,
and \lgg\,=\,4.0 was adopted. The vertical bars show position of the \ion{Ca}{ii} IR triplet
lines.}
\label{hbop}
\end{figure}

\txtbf{In all spectrum synthesis calculations in this paper we use zero microturbulence because the
{\tt SYNTHMAG} code models magnetic intensification by directly including magnetic field effects in
the solution of the polarized radiative transfer equation. The true microturbulence is thought be 
absent in Ap stars because kG-strength magnetic field is quite efficient in suppressing
convective motions.}

\txtbf{Here we analyse the target stars assuming a homogeneous distribution of Ca over their
surfaces. This assumptions appears to be reasonable given a slow rotation of the majority of the
targets. Moreover, no evidence of substantial horizontal
Ca abundance gradients exist for any of the studied stars, although many of them were observed at high
resolution more than once (e.g., Mathys et al. \cite{mathys97}; Ryabchikova et al. \cite{RNW04}).
The only star from our sample for which the Ca surface inhomogeneity was investigated
previously, is HD\,24712 (L\"uftinger et al. \cite{LKR07}). The full range of the horizontal 
Ca abundance variation in this star is only 0.2~dex. The resulting effect on Ca the line profiles
is negligible compared to the vertical abundance jump of more than 3~dex (see below). We also note
that, even if hypothetical high-contrast Ca spots exist on the surface of some of our stars, their
effect will be to weight our modelling results to specific surface areas. But none of the potential 
horizontal inhomogeneity effects can possibly mimic stratification or isotopic anomaly studied here.}

An important modification to the {\tt SYNTHMAG} treatment of the hydrogen line opacity was
introduced to improve analysis of the IR \ion{Ca}{ii} triplet lines. These \ion{Ca}{ii}
transitions blend with the high members of the Paschen series, which become prominent in Ap stars
with  \teff\,$\ge$\,8000~K. The usual calculation of the overlapping hydrogen line wings, which
adds linearly opacities due to bound-free and bound-bound hydrogen opacity,  becomes increasingly
inaccurate as one approaches the series limit. As illustrated in Fig.~\ref{hbop}, this leads to a
spurious reduction in calculated fluxes. The correct treatment of the overlapping hydrogen line
and continuous opacity should be done according to the occupation probability formalism (Daeppen
et al. \cite{DAM87}; Hubeny et al. \cite{HHL94}). This technique was implemented in {\tt SYNTHMAG}
with the help of {\tt HBOP} procedure, developed by P. Barklem at Uppsala Observatory\footnote{\tt
http://www.astro.uu.se/\~{}barklem/hlinop.html}. This hydrogen opacity code, which also
incorporates previous developments in the hydrogen line broadening theory 
(Stehl\'e \& Hutcheon \cite{SH99}; Barklem et al.
\cite{BPO00}), was applied for the spectrum synthesis of all stars in our sample. Fig.~\ref{hbop}
shows that correct treatment of the hydrogen opacity significantly reduces predicted Paschen line
absorption at the position of the \ion{Ca}{ii} 8498\,\AA\ line analysed here.

Magnetic spectrum synthesis with {\tt SYNTHMAG} was coupled with the vertical abundance mapping
procedure {\tt DDAFIT}, described by Kochukhov (\cite{synthmag06}) and used previously by
Ryabchikova et al. (\cite{RLK05,RRKB06}). This routine provides a graphical and optimization
interface to {\tt SYNTHMAG}, allowing the user to find parameters of a simple stratification
model by fitting a large number of spectral lines of a given element.

\section{Ca stratification analysis}
\label{strat}

\begin{figure*}[!th]
\centering
\resizebox{15cm}{!}{\includegraphics{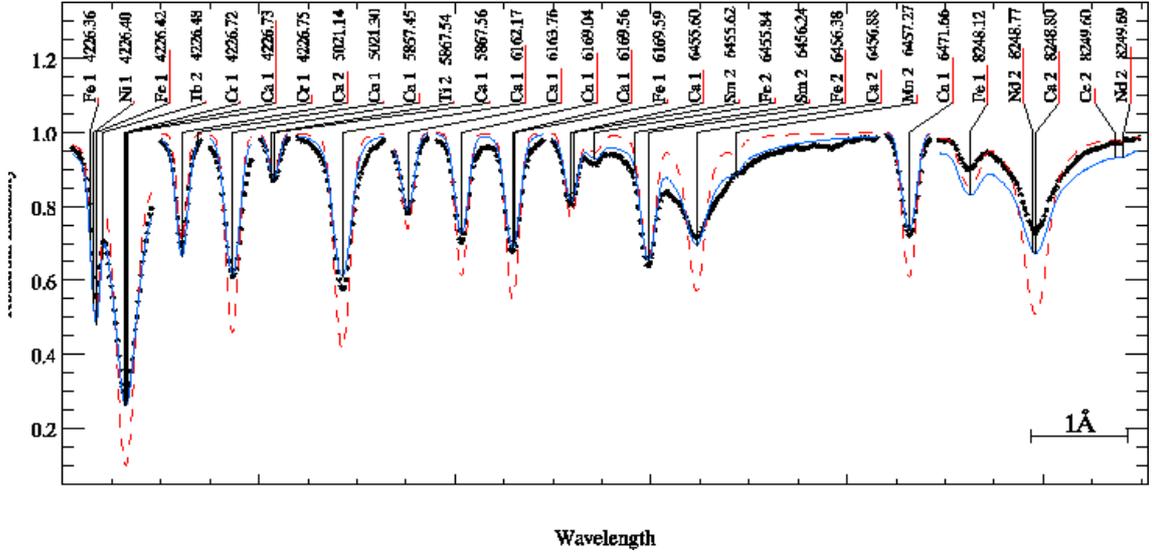}}
\caption{Comparison between the observed line profiles (symbols) of HD~176232 
(10\,Aql) and theoretical 
spectra calculated with the best-fitting stratified Ca abundance distribution (solid line)
and with the homogeneous Ca abundance (dashed line).
The plotted spectral windows show segments of the HD~176232 spectra used for determination of
Ca stratification; symbols show actual observed points.
}
\label{10_Aql_Ca}
\end{figure*}

Before performing detailed study of the IR \ion{Ca}{ii} $\lambda$~8498\,\AA\
line profile, one has to derive vertical distribution of Ca abundance in the 
atmospheres of Ap stars. Without this step, a quantitative analysis of isotopic
anomaly is practically impossible because profiles of all strong Ca lines,
including the IR triplet, are dramatically affected by the vertical
stratification of this element. In all program stars Ca stratification was
derived using a set of spectral lines in the optical region, for which no
indication on the significant isotopic shifts exists. 
\txtbf{In this way we decouple stratification analysis from the study
of isotopic shifts. At this stage we also avoid using the \ion{Ca}{ii} 8498\,\AA\ line
because of the possible sensitivity to the errors in effective temperature
due to the special normalization procedure adopted for this line (see Sect.~2).}

Atomic parameters of \txtbf{the Ca lines} are given in
Table\,\ref{strat-list}. Stratification analysis requires high accuracy not
only for the oscillator strengths but also for the damping parameters, because
Ca has a tendency to be concentrated close to the photospheric layers where the
electron density is high.  Accurate treatment of damping wings is particularly
important for the lines of ionized Ca. For the \ion{Ca}{ii} lines at
$\lambda\lambda$ 3158, 3933, 8248, 8254, 8498, 8912, 8927\,\AA\ the Stark damping constants
were taken from the paper by Dimitrijevi\'c \& Sahal-Br\'echot (\cite{Stark}),
where semi-classical calculations as well as a compilation of the experimental
data was presented. For the rest of Ca lines the Stark damping constants
calculated by Kurucz (\cite{K93}) were adopted. The oscillator strengths were
taken mostly from the laboratory experiments -- these data were  thoroughly
verified by the recent NLTE analysis of calcium in late-type stars (Mashonkina
et al. \cite{MKP07}). Because of the large range in effective temperatures and
magnetic field strengths, we could not use the same set of lines for all
stars. 

\begin{table}[!t]
\caption{Atomic parameters of the spectral lines employed in the Ca
stratification analysis. The columns give the ion designation,
central wavelength of the transition, the excitation potential of the lower level, 
oscillator strength, the Stark damping constant, and the reference for 
the oscillator strength. \label{strat-list}}
\begin{center}
\begin{tabular}{lcrrcc}
\hline
\hline
Ion &$\lambda$ (\AA) &\ei\,(eV)  &$\log\,{gf}$&$\log\,\gamma_{\rm St}$ & Reference \\
\hline
\ion{Ca}{ii}&  3158.869& 3.123 &    0.241& $-$4.90 & T  \\ 
\ion{Ca}{ii}&  3933.655& 0.000 &    0.105& $-$5.73 & T  \\ 
\ion{Ca}{i} &  4226.728& 0.000 &    0.244& $-$6.03 & SG \\ 
\ion{Ca}{ii}&  5021.138& 7.515 & $-$1.207& $-$4.61 & SMP \\ 
\ion{Ca}{ii}&  5339.188& 8.438 & $-$0.079& $-$3.70 & SMP \\
\ion{Ca}{i} &  5857.451& 2.933 &    0.240& $-$5.42 & S \\ 
\ion{Ca}{i} &  5867.562& 2.933 & $-$1.570& $-$4.70 & S \\ 
\ion{Ca}{i} &  6122.217& 1.896 & $-$0.316& $-$5.32 & SO \\ 
\ion{Ca}{i} &  6162.173& 1.899 & $-$0.090& $-$5.32 & SO \\ 
\ion{Ca}{i} &  6163.755& 2.521 & $-$1.286& $-$5.00 & SR \\ 
\ion{Ca}{i} &  6166.439& 2.521 & $-$1.142& $-$5.00 & SR \\ 
\ion{Ca}{i} &  6169.042& 2.253 & $-$0.797& $-$5.00 & SR \\ 
\ion{Ca}{i} &  6169.563& 2.256 & $-$0.478& $-$4.99 & SR \\ 
\ion{Ca}{i} &  6449.808& 2.521 & $-$0.502& $-$6.07 & SR \\ 
\ion{Ca}{i} &  6455.598& 2.523 & $-$1.340& $-$6.07 & S \\ 
\ion{Ca}{ii}&  6456.875& 8.438 &    0.410& $-$3.70 & SMP \\ 
\ion{Ca}{i} &  6462.567& 2.523 &    0.262& $-$6.07 & SR \\ 
\ion{Ca}{i} &  6471.662& 2.526 & $-$0.686& $-$6.07 & SR \\ 
\ion{Ca}{ii}&  8248.796& 7.515 &    0.556& $-$4.60 & SMP \\ 
\ion{Ca}{ii}&  8254.721& 7.515 & $-$0.398& $-$4.60 & SMP \\ 
\ion{Ca}{ii}&  8498.023& 1.692 & $-$1.416& $-$5.70 & T  \\ 
\ion{Ca}{ii}&  8912.068& 7.047 &    0.637& $-$5.10 & SMP\\ 
\ion{Ca}{ii}&  8927.356& 7.050 &    0.811& $-$5.10 & SMP\\ 
\hline
\end{tabular}
\end{center}
Rerefences. T -- Teodosiou (\cite{T}); SG -- Smith \& Gallagher (\cite{SG}); SMP -- Seaton et al. (\cite{TB});
S -- Smith (\cite{S}); SO -- Smith \& O'Neil (\cite{SN}); SR -- Smith \& Raggett (\cite{SR}).
\end{table}

The Ca stratification analysis was performed using the step-function
approximation of the abundance distribution (for details see, e.g., Ryabchikova
et al. \cite{RLK05}). This parameterized model of the vertical stratification
is characterized by the upper and lower abundance values, as well as by the
position and width of abundance jump. In each star these parameters are
constrained by simultaneous fit to the profiles of many lines. These observational data
are sufficient to constrain stratification parameters with a good precision.

\begin{figure}[!t]
\centering
\fifps{90mm}{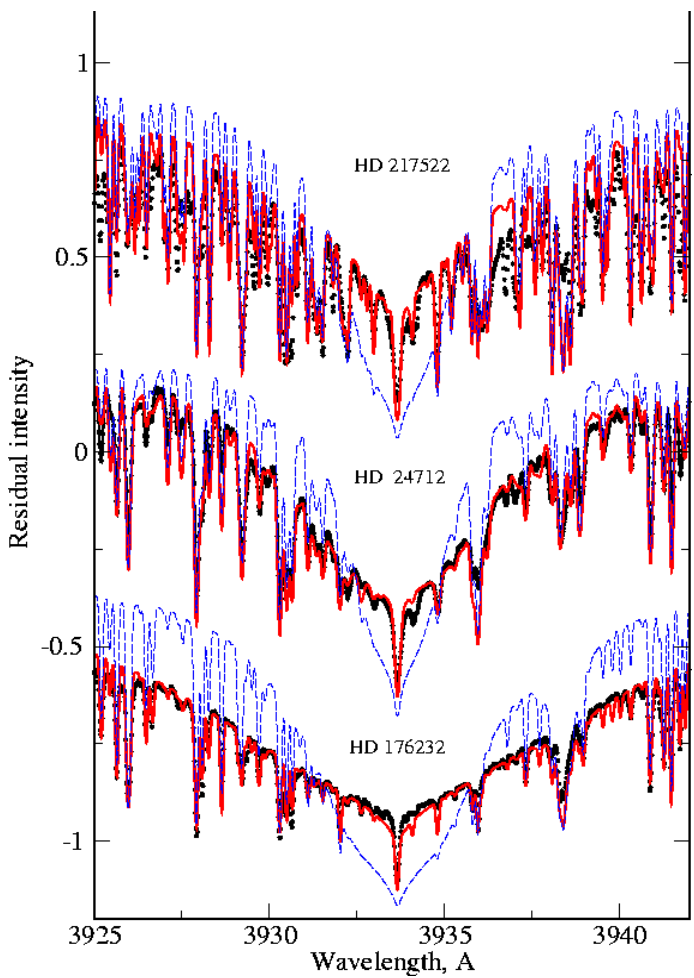}
\caption{Comparison between the observed profiles (symbols) of the \ion{Ca}{ii} 3933\,\AA\
line and calculations with the stratified (solid line) and homogeneous (dashed line)
Ca distributions for 3 program stars. The spectra of HD\,24712 and HD\,176232 are shifted
downwards for display purpose.}
\label{Ca2_3933}
\end{figure}

\begin{figure}[!t]
\figps{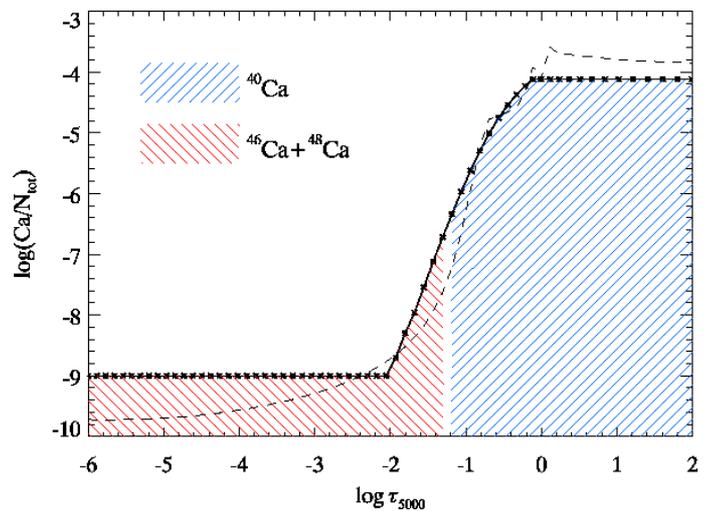}
\caption{Vertical stratification and isotopic separation of Ca derived for
HD\,176232. The overall Ca abundance profile, inferred from the stratification
analysis of optical lines, is shown with the solid line. The hatched areas demonstrate
vertical separation of the light and heavy Ca isotopes required to fit the \ion{Ca}{ii}
8498\,\AA\ line. \txtbf{The dashed line shows the Ca distribution derived the with VIP code
using the same set of lines.}}
\label{ca_strat}
\end{figure}

We start stratification analysis of Ca by computing the optimal  homogeneous
Ca abundance for a chosen set of spectral lines. Then we let the {\tt DDAFIT}
procedure to vary parameters of the step-function until the adequate fit to the
observed line profiles is achieved. The outcome of this procedure is illustrated in 
Fig.~\ref{10_Aql_Ca}, where results of the stratification analysis for the set
of optical lines in HD~176232 (10~Aql) are displayed. 
Figure~\ref{Ca2_3933} shows that stratified Ca distributions obtained for 
HD~176232 and other cool Ap stars also dramatically improve the fit to 
the strongest \ion{Ca}{ii} 3933\,\AA\ line compared to the spectrum synthesis with a
homogeneous Ca distribution. In Figs.~\ref{10_Aql_Ca} and \ref{Ca2_3933}
synthetic profiles
calculated with the uniform Ca distribution $\log (Ca/N_{\rm tot})=-5.14$ are
shown by dashed lines while those calculated with the best-fitting stratified Ca
distribution are shown by the solid lines. Clearly, even a very schematic Ca
stratification model used here provides a substantial improvement of the 
agreement between observations and theoretical calculations.
We use standard deviation to characterize the quality of the fit. This 
parameter is defined as a rms of the observed minus calculated spectrum, evaluated
on the wavelength grid of observations within the spectral windows covering
lines of interest (see Fig.~\ref{10_Aql_Ca}). For HD~176232 and most other stars
the stratified Ca abundance yields roughly two times smaller
standard deviation compared to the homogeneous Ca distribution. 
The final
Ca stratification model inferred for HD~176232 is illustrated in Fig.~\ref{ca_strat}.

To verify that the step-function approximation provides a realistic description of the Ca
distribution we performed for a few stars with weak magnetic fields 
(HD~133792, HD~176232, HD~203932)
stratification analysis using the Vertical Inverse Problem code, {\tt VIP}. This program 
does not use any \textit{a priori} assumptions about
the shape of the vertical element abundance profile but currently cannot be applied to stars
with a strong magnetic field
(Kochukhov et al. \cite{KTRM06}). 
A comparison between the Ca distributions derived for HD~176232 with {\tt DDAFIT}
and with {\tt VIP} is shown in Fig.~\ref{ca_strat}. The same level of agreement
was obtained for the two other stars.    

However, despite general success of the {\tt DDAFIT} modelling,  
in a few stars the step-function
approximation could not provide an adequate description of the full set of
spectral lines. The obvious reason is the use of normal non-magnetic stellar model
atmosphere with a homogeneous element distribution for a star with abundance
stratification, and a deviation of the Ca abundance distribution from the
simple step-function form. 

In cooler stars we do not systematically use the strongest \ion{Ca}{ii} 3933\,\AA\
line  in the stratification analysis due to difficulties of treating blends in the extended
wings and establishing continuum. The range of the formation depths of the remaining weaker
optical lines is different from the IR triplet lines of interest, therefore Ca abundance in
the upper atmospheric layers is poorly constrained by the optical lines and may not be
accurate enough for the description of the cores of the IR \ion{Ca}{ii} triplet lines. 
This is often reflected by a rather large formal error of the abundance in the upper layers 
compared with
the corresponding error of the abundance in the lower layers. For several stars 
(see Table \,\ref{ca_strat_tbl}) a change of the  Ca abundance above in the upper layers 
within
3$\sigma$ was introduced to fit overall intensity of the IR triplet line cores. 
In addition for a few other stars further
improvement of the fit quality was possible by introducing  a more complicated stratification
profile starting from the results obtained with {\tt DDAFIT}. This step was performed
manually for 5 out of 23 Ap stars. In all the cases when we have modified Ca stratification
determined with {\tt DDAFIT}, the consistency of the altered vertical abundance profile with
all Ca diagnostic lines was verified. 

The formal stratification analysis of the Am star HD\,27411 (whose \teff\
is identical to that of HD\,176232) and normal star HD\,73666 (\teff=9382 K) yields a 0.2~dex abundance jump, 
which we consider to be insignificant. 
The average Ca abundance 
$\log ({\rm Ca}/N_{\rm tot})=-5.63$ 
in both stars represents very well most of the lines, including the resonance \ion{Ca}{ii} 3933\,\AA\ line. 
In both stars 0.1~dex lower abundance is required to fit the  \ion{Ca}{ii} 3933\,\AA\ profile compared to
\ion{Ca}{ii} 8498\,\AA\ line.
This internal agreement is excellent considering difficulties in
continuum normalization. It confirms that the Ca abundance variation by several
dex over the vertical span of the Ap-star atmospheres is real.

\section{The Ca isotopic anomaly}
\label{isot}

Ca has six stable isotopes with atomic numbers 40, 42, 43, 44, 46, 48.  In the terrestrial matter
Ca mixture consists mainly of $^{40}$Ca (96.9 \% according Anders \& Grevesse \cite{AG89}).
Table\,\ref{iso8498} gives wavelengths of all Ca isotopes in the \ion{Ca}{ii} 8498\,\AA\
transition based on the isotopic shifts measured by N\"ortersh\"auser et al. (\cite{Ca_IS}).
We also list fractional oscillator strengths corresponding to the terrestrial isotopic
mixture. 

Isotopic shifts were also measured for the \ion{Ca}{ii} resonance lines at $\lambda\lambda$ 
3933 and 3968\,\AA\ (M{\aa}rtensson-Pendrill et al. \cite{MYW92}). 
The wavelength shift between the $^{48}$Ca and 
$^{40}$Ca isotopes is $-0.009$\,\AA, and can be observed only for the \ion{Ca}{ii} 3933\,\AA\ line 
core in hot stars with an especially large contribution of the $^{48}$Ca isotope. 
One star from our sample, HD~133792, shows an isotopic shift in the \ion{Ca}{ii} 3933\,\AA\ line.

\begin{table}
\caption{Atomic data for the isotopic components of \ion{Ca}{ii}
$\lambda$~8498\,\AA. The fractional abundance of Ca isotopes, $\varepsilon$,
corresponds to the composition of the terrestrial matter.
\label{iso8498}}
\begin{center}
\begin{tabular}{ccc}
\hline 
\hline 
 \ $\lambda$ (\AA) & isotope  & $\log gf\varepsilon$ \\
\hline
  8498.023 & 40 &  $-$1.43   \\
  8498.079 & 42 &  $-$3.60   \\
  8498.106 & 43 &  $-$4.29   \\
  8498.131 & 44 &  $-$3.10   \\
  8498.179 & 46 &  $-$5.81   \\
  8498.223 & 48 &  $-$4.14   \\
\hline
\end{tabular}
\end{center}
\end{table}

\begin{figure}[!t]
\centering
\fifps{90mm}{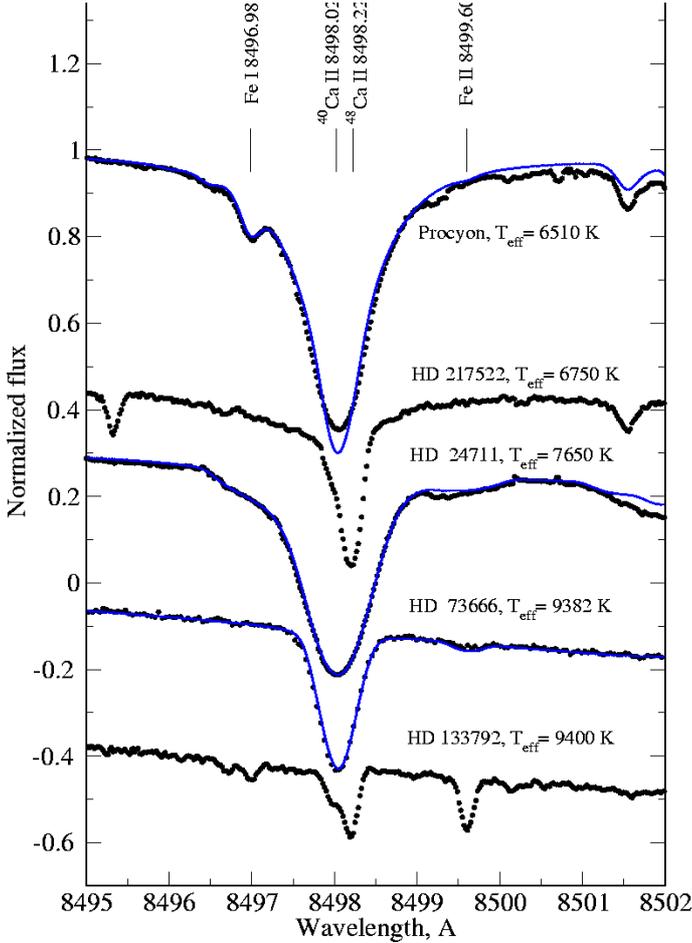}
\caption{Comparison between the observed line profiles (symbols) of the \ion{Ca}{ii} 8498\,\AA\
line and calculations for the normal Ca isotopic mixture (solid line) in the spectra of Procyon, 
Am star HD~27411, and A1 V star HD~73666. The observed spectrum of the cool (HD~217522) and tepid (HD~133792) 
Ap stars, presented in the 
middle and bottom, clearly shows a displaced \ion{Ca}{ii} line core. The spectra 
are shifted downwards for display purpose.}
\label{Ca_ref}
\end{figure}

Using the terrestrial isotopic mixture, we calculated the profile of the  \ion{Ca}{ii} 8498 line
in the spectra of our reference stars, Procyon, HD~27411 and HD~73666. Comparison of the theoretical
spectra and observations is illustrated in Fig.~\ref{Ca_ref}. Although in the Procyon spectrum
our LTE calculations cannot provide a very good fit to the deepest part of the core, no
wavelength shifts are detected in either star. At the same time, the observed profile of the
\ion{Ca}{ii} line in the spectrum of one of our program stars, HD~217522, presented in
Fig.~\ref{Ca_ref} exhibits a complex structure and is clearly redshifted, with the strongest
component coinciding in wavelength with the expected position of the heaviest Ca isotope. This is
a signature of the Ca isotopic anomaly and, as we will show below, of the vertical separation of
Ca isotopes.

\begin{figure}[!t]
\centering
\figps{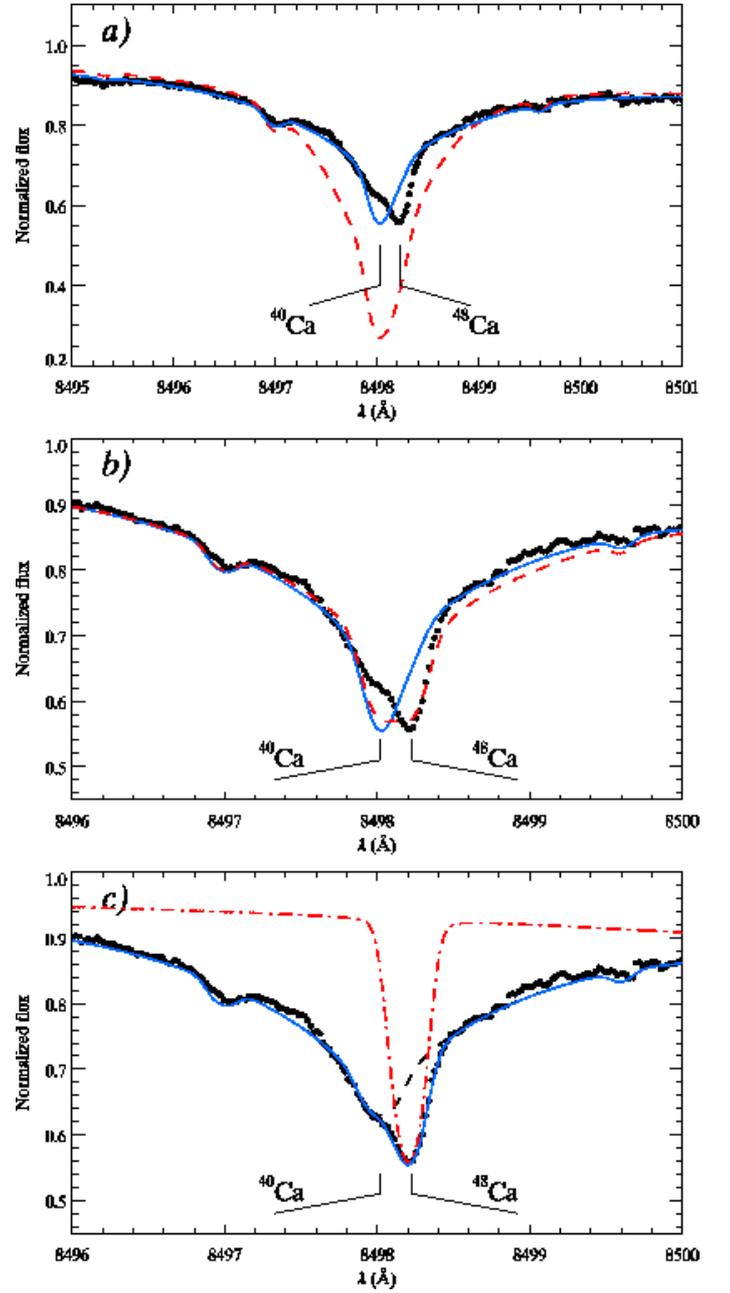}
\caption{
Derivation of the isotopic composition and its height dependence for
the \ion{Ca}{ii} 8498\,\AA\ line in HD\,176232 (10\,Aql). The three panels show
the UVES observations (full circles) compared to various modelling attempts, as 
follows. 
{\bf a)} 
Dashed line: theoretical spectrum obtained with a homogeneous Ca distribution.
Solid line:  theoretical spectrum calculated for a stratified Ca abundance.
A terrestrial mixture of Ca isotopes is adopted for these computations. 
{\bf b)}  
Solid line: theoretical spectrum obtained for a terrestrial Ca isotopic composition (same
as panel a); 
Dashed line: theoretical spectrum obtained for 
a 50:50 ratio of $^{40}$Ca to $^{46}$Ca+$^{48}$Ca. 
{\bf c)} 
Solid line: the \ion{Ca}{ii} line profile computed for the best-fitting model of vertical
isotopic separation (see Fig.~\ref{ca_strat}).
Dashed and dash-dotted lines show contributions of $^{40}$Ca and $^{46}$Ca+$^{48}$Ca, respectively.
The wavelength position of $^{40}$Ca and $^{48}$Ca is indicated in each panel.
These plots show that an acceptable fit to the observations can only be achieved
with a depth-dependent Ca isotopic mixture.
}
\label{HD176232_8498}
\end{figure}

The core of the IR \ion{Ca}{ii} 8498\,\AA\ line is formed higher than any of the optical
lines, except \ion{Ca}{ii} 3933\,\AA. For most stars the \ion{Ca}{ii}
3933\,\AA\ line was not modelled in our stratification calculations. Therefore, Ca
abundance in the upper atmosphere of some stars may be somewhat uncertain, as all other optical lines are not
sensitive to the abundance variations above $\log\tau_{5000}=-2.0$ to $-2.5$. 
However, for the majority of our targets the Ca abundance in the
upper atmosphere, responsible the strength of the \ion{Ca}{ii} 8498\,\AA\ line core, is
defined by the slope of the abundance gradient in the jump region. If the Ap atmospheric
structure is close to the normal {\tt ATLAS9} atmosphere adopted in our analysis, then the
\ion{Ca}{ii} 8498\,\AA\ line should be well-described by the Ca abundance distribution
derived from optical lines. Our calculations show that while it is correct for the observed total
intensity, in many of the program stars we cannot fit the line cores, which are often redshifted.
This situation is illustrated in Fig.~\ref{HD176232_8498}a, where we compare the observed
spectrum of HD~176232 with the synthetic spectrum calculated for the terrestrial Ca isotopic
mixture and Ca abundance distribution (Fig.~\ref{ca_strat}) derived from optical lines. One
immediately notices that while the line wings are explained by our calculations, the line core
cannot be fitted with the terrestrial Ca isotopic mixture.

Cowley \& Hubrig (\cite{CH05}) showed that the red-shift of the Ca IR line core arises due to 
the heavy Ca isotopes and claimed that ``in extreme cases the dominant isotope is the exotic 
$^{48}$Ca''.
A simple interpretation of the anomaly observed in the \ion{Ca}{ii} 8498\,\AA\ line core is to
suggest that heavy Ca isotopes are strongly enhanced and even dominant throughout the atmospheres
of some magnetic Ap stars. 
However, our magnetic spectrum synthesis calculations demonstrate that this
hypothesis is incorrect. On the example of HD\,176232 we can show that using a model with 
stratified Ca distribution
found previously and assuming a depth-independent enhancement of $^{46}$Ca and $^{48}$Ca relative
to $^{40}$Ca, the center-of-gravity of the \ion{Ca}{ii} 8498\,\AA\ line core is shifted to the red, but
the overall fit to the line profile does not improve considerably (Fig.~\ref{HD176232_8498}b). In
disagreement with observations, a constant overabundance of the heavy Ca isotopes gives rise to
the red-shifted long-wavelength wing of the \ion{Ca}{ii} line.

Here we propose a different explanation of the \ion{Ca}{ii} 8498\,\AA\ line shape. Observations
of HD\,176232 and other stars with displaced Ca line core show a shallow line with wide wings
at  the position of $^{40}$Ca isotope, which means that this line component is formed in the
lower atmospheric layers. On the other hand, at the position of $^{48}$Ca we see a sharp deep
line, which is responsible for a characteristic steep intensity gradient in the red wing of the
\ion{Ca}{ii} 8498\,\AA\ line. We observe the same picture in all cool Ap stars with significant
isotopic shifts of the  \ion{Ca}{ii} line core. Given the Ca stratification inferred for all Ap
stars, this deep component lacking developed wings can be explained by the absorption in
the upper atmosphere, above the Ca abundance jump or in the outermost part of the transition
region. Thus, observations of the \ion{Ca}{ii} 8498\,\AA\ line in Ap stars should be interpreted
in terms of \textit{vertical separation of Ca isotopes in the stellar atmosphere}.

If we separate $^{40}$Ca and the sum of $^{46}$Ca and $^{48}$Ca isotopes in the atmosphere of
HD\,176232 as indicated in Fig.~\ref{ca_strat} (the boundary between heavy and light Ca is at 
$\log\tau_{\rm 5000}=-1.2$), a satisfactory agreement between the observed and calculated
spectra is obtained (see Fig.~\ref{HD176232_8498}c).  The dominant atmospheric constituent is
still the normal isotope $^{40}$Ca, but the shape of the \ion{Ca}{ii} 8498\,\AA\ line core is
influenced by the high-lying cloud of heavy Ca, in which $^{46}$Ca and $^{48}$Ca contribute 
35\% and 65\%, respectively.

Thus, in our modelling of the Ca isotopic anomaly in Ap stars we adopt an initial model in
which the total element abundance is given by the stratification profile obtained
previously, while the vertical separation of the light and heavy Ca isotopes occurs nearly
instantaneously and is characterized by a single transition depth, deduced from the
\ion{Ca}{ii} 8498\,\AA\ line. Relative contributions of $^{46}$Ca and $^{48}$Ca to the heavy
Ca layer in the upper atmosphere are also adjusted based on the \ion{Ca}{ii} IR triplet
line. 

\begin{figure}[!th]
\centering
\fifps{90mm}{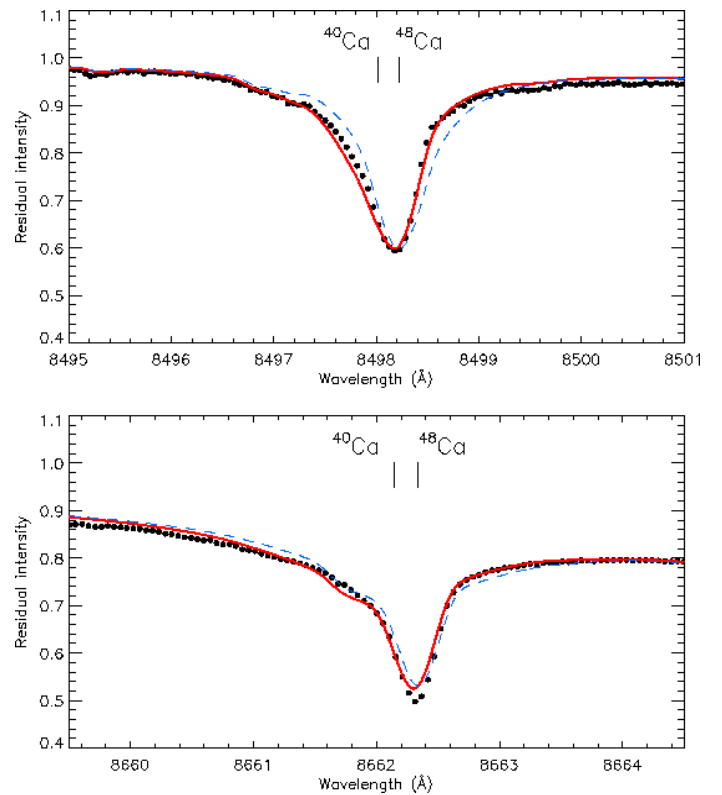}
\caption{Comparison of the observed IR triplet lines (symbols) in the
ESPaDOnS spectrum of HD~24712 and theoretical spectrum synthesis for
the best-fitting stratified Ca abundance distribution (solid line)
with Ca isotopic separation and with the stratified distribution of $^{48}$Ca only
(dashed line). }
\label{HD24712_IR}
\end{figure}
 
Parameterization of the vertical separation of Ca isotopes that we adopt here is just 
one of several models possible. For instance, we find that an acceptable fit to the
\ion{Ca}{ii} 8498\,\AA\ line in some of the target stars could be obtained with a model
where $^{48}$Ca is homogeneously distributed while $^{40}$Ca is stratified. (Such a model is
achieved by placing the border between Ca isotopes horizontally rather than vertically in
Fig.~\ref{ca_strat}. Note, that formally $^{48}$Ca is not anomalous: its relative-to-hydrogen abundance 
in all stars does not exceed an abundance in the  normal solar Ca mixture.) This model turns out to be nearly equivalent to the one we use because
the boundary between the heavy and light Ca layers is usually located very close to
the position of the Ca abundance jump. The $^{40}$Ca contribution still dominates in the
lower atmosphere, whereas $^{48}$Ca contributes to the light absorption in the upper layers.
However, for several stars where we need Ca enhancement in the uppermost layers, the model with
homogeneously distributed heavy Ca isotopes clearly yields worse agreement with observations
in comparison with the vertical separation model adopted in here.

Our isotopic stratification model is based upon only one of the IR triplet lines. 
High-quality observations of the two other IR triplet lines are unavailable for the majority of 
our targets. For one of the cool Ap stars, HD\,24712, we assessed the quality of the
fit to all three \ion{Ca}{ii} lines using the spectrum obtained with ESPaDOnS at CFHT.
These data were collected at the rotational phase different from that of UVES observations, 
therefore we have performed an independent Ca stratification and isotopic analysis for HD~24712
based on the ESPaDOnS spectrum alone.
Fig.~\ref{HD24712_IR} compares calculated and observed \ion{Ca}{ii} 8498\,\AA\ 
and 8662\,\AA\ line profiles. Due to specific position of the \ion{Ca}{ii} 8542\,\AA\ line 
near the beginning of
the spectral order, which makes continuum normalization uncertain, we did not include this line to
the plot.

Of course, the proposed model of Ca isotopic separation is at best a crude approximation of
potentially much more complex isotopic stratifications. However, we find that, for many
stars, introducing a combination  of stratification and isotopic segregation noticeably
improves the fit to the IR triplet lines and thus provides a direct evidence for the Ca
isotopic separation in the atmospheres of Ap stars. 

\txtbf{Finally, we note that the continuum normalization in the region
around IRT lines is indirectly based on the adopted effective temperatures, which may
introduce an additional uncertainty and sometimes lead to a poor fit in the line
wings. However, we have verified that even considerable 
errors in \teff\ lead only to a small 
change of the overall depth of the \ion{Ca}{ii} 8498\,\AA\ line computed for a given Ca
stratification. Since this line is located
far enough from the H line core, its profile, especially the shape of the inner core region used 
in the isotopic analysis, is generally not affected by the \teff\ errors.
}

\begin{figure}[!t]
\figps{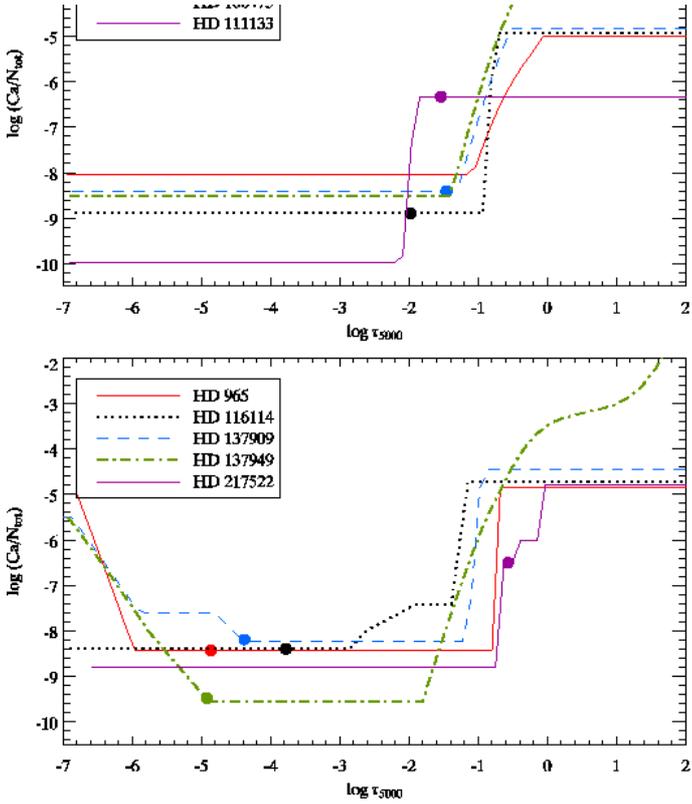}
\caption{Ca stratification in the atmospheres of selected program stars. Upper panel: distribution of Ca
in HD\,66318, HD\,75445, HD\,134214, HD\,166473 and HD\,111133. For stars enriched in heavy Ca
the filled circle indicates the bottom of the heavy Ca isotope layer. Lower panel: same for stars with
complex distribution of Ca abundance with height (HD\,965, HD\,116114, HD\,137909, HD\,137949, HD\,217522).
}
\label{Ca_distr}
\end{figure}

\section{Results}
\label{results}

\begin{table*}
\caption{Parameters of the Ca stratification and isotopic anomaly determined for magnetic CP stars. The
columns give HD number of stars, step-function parameters of the vertical Ca distributions 
(abundance in upper atmosphere, abundance in the lower layers, position of the abundance
jump and the width of the transition zone), parameters of the Ca isotopic separation if applicable (the
optical depth of the boundary between the heavy and light Ca layers, the 
dominant isotopic component of
 the heavy Ca cloud). The last column indicates deviation
of the final Ca stratification profile from a simple step-function.
\label{ca_strat_tbl}}
\begin{center}
\begin{tabular}{r|cccc|cc|c}
\hline 
\hline 
 HD~~~  & \multicolumn{4}{c|}{Stratification parameters}   
        & \multicolumn{2}{c|}{Ca isotopic anomaly} & Comment  \\
 number & $\log (Ca/N_{\rm tot})_{\rm up}$ & $\log (Ca/N_{\rm tot})_{\rm lo}$
        & $\log\tau_{\rm 5000}(step)$ & $\Delta\log\tau_{\rm 5000}(step)$
        & $\log\tau_{\rm 5000}(isot)$ & isotope & \\
\hline
   965 & ~$-8.4$ & $-4.9$ & $-0.7$ & 0.1 & $-4.9$ & $^{46}$Ca      & $a,d$ \\   
 24712 & ~$-8.7$ & $-5.1$ & $-1.2$ & 2.5 & $-1.4$ & $^{46}$Ca      & $c,d$ \\   
 29578 & ~$-8.9$ & $-4.3$ & $-1.2$ & 0.7 &         &                &  \\      
 47103 & ~$-7.7$ & $-4.6$ & $-0.8$ & 0.3 &         &                &  \\      
 66318 & ~$-8.0$ & $-5.0$ & $-0.6$ & 1.1 &         &                &  \\      
 75445 & ~$-8.9$ & $-4.9$ & $-0.9$ & 0.2 & $-2.0$ & $^{48}$Ca      &  \\      
101065 & ~$-8.9$ & $-5.3$ & $-0.4$ & 0.4 & $-0.2$ & $^{48}$Ca      &  \\      
111133 & $-10.0$ & $-6.3$ & $-2.0$ & 0.4 & $-1.5$ & $^{44}$Ca      &  \\      
116114 & ~$-7.4$ & $-4.7$ & $-1.3$ & 0.2 & $-3.8$ & $^{48}$Ca      & $b$ \\   
118022 & ~$-7.0$ & $-2.0$ & $-0.1$ & 0.3 & $-1.5$ & $^{48}$Ca      &  \\      
122970 & ~$-7.7$ & $-5.1$ & $-1.5$ & 2.8 & $-1.6$ & $^{46}$Ca      & $c,d$ \\   
128898 & ~$-8.5$ & $-4.0$ & $-1.0$ & 2.0 & $-1.9$ & $^{48}$Ca      & $c,d$ \\   
133792 & ~$-8.1$ & $-5.6$ & $-0.6$ & 0.1 & $-1.0$ & $^{48}$Ca      &  \\      
134214 & ~$-8.4$ & $-4.8$ & $-0.9$ & 0.9 & $-1.5$ & $^{48}$Ca      &  \\      
137909 & ~$-8.2$ & $-4.4$ & $-1.0$ & 0.4 & $-4.4$ & $^{48}$Ca      & $a$ \\   
137949 & ~$-9.6$ & $-3.5$ & $-1.0$ & 1.8 & $-4.9$ & $^{48}$Ca      & $a,c$ \\  
144897 & ~$-8.5$ & $-5.2$ & $-1.9$  & 0.5  &         &                &  \\     
166473 & ~$-8.5$ & $-3.8$ & $-1.0$ & 1.1 &         &                &  $d$  \\       
170973 & ~$-5.3$ & $-5.0$ & $-1.1$ & 0.1 &         &                &  \\       
176232 & ~$-9.0$ & $-4.1$ & $-1.2$ & 1.8 & $-1.2$ & $^{48}$Ca      &   $d$ \\       
188041 & ~$-7.4$ & $-3.1$ & $-1.2$ & 2.5 & $-3.7$ & $^{46}$Ca      & $c,d$ \\  
203932 & ~$-8.7$ & $-4.5$ & $-1.2$ & 2.6 & $-2.1$ & $^{48}$Ca      & $c,d$ \\    
217522 & ~$-8.8$ & $-4.8$ & $-0.3$ & 0.7 & $-0.6$ & $^{48}$Ca      & $b$ \\    
\hline               
\end{tabular}
\end{center}
Comments. $a$ -- abundance increase in 
the upper layers; $b$ -- complex transition zone; $c$ -- distortion of the step-function after 
$\log \rho x$ to $\log\tau_{\rm 5000}$ transformation; $d$ -- a change in the upper abundance value.
\end{table*}

\subsection{Ca stratification}

The Ca stratification analysis described in Sect.~\ref{strat} was applied  to
all Ap stars included in our sample. The presence of a stratified Ca distribution
is inferred for all stars except the hot Ap star HD\,170973,
where stratification appears to be negligible. 
Table~\ref{ca_strat_tbl} gives parameters of the
step-function model distributions of Ca abundance. These vertical abundance profiles are
characterized by an abundance jump in the atmospheric region
$-1.3\le\log\tau_{5000}\le-0.5$, a 1--1.5 dex overabundance deep in the atmosphere and a
strong Ca depletion above $\log\tau_{5000}\approx-1.5$. 
\txtbf{The formal uncertainty of the stratification parameters is typically 0.3--0.7~dex, 
0.1--0.3~dex and $<$\,0.1~dex for
the upper abundance, lower abundance and the position of the Ca abundance jump, respectively.}
The Ca stratification in the
atmospheres of selected program stars is shown in Fig.~\ref{Ca_distr}. 

For several stars a more complex vertical distributions of Ca abundance had to be introduced.
Observations of HD\,965, HD\,137909, HD\,137949 are better reproduced with an increase of Ca
concentration in the upper atmospheric layers above $\log\tau_{5000}=-5$. Although seemingly
complicated, this picture does not contradict theoretical Ca diffusion calculations. Both
Borsenberger et al. (\cite{BPM81}, Fig.~6) and Babel (\cite{Babel92}) obtained Ca abundance
increase in the uppermost layers, above the main abundance jump. However, NLTE treatment of the Ca lines
formation is needed to investigate stratification of Ca in the upper atmospheric layers in more
detail. For two other stars, HD\,116114 and HD\,217522, a structure of the transition zone 
somewhat more complex than a linear transition between two constant values is required to fit
observations. Figure~\ref{Ca_distr} (lower panel) illustrates the final chemical stratification 
profiles adopted for the five stars where Ca stratification deviates from a step-function profile.

We note that for all stars the {\tt DDAFIT} inversion was performed on the original
column mass ($\log \rho x$) depth grids of the {\tt ATLAS9} models. 
Table~\ref{ca_strat_tbl} gives vertical stratification  parameters converted to the
$\log\tau_{5000}$ scale. In the cases when transition zone is wide, extending to the
lower atmospheric layers, transformation between
the column mass and the standard optical depth scale becomes non-linear, which leads to a
distortion of the step-function shape (e.g., Ca distribution for HD\,137949 in 
Fig.~\ref{Ca_distr}). 
\txtbf{For the same reason in some of the stars where this problem was encountered  (e.g.,
HD~137949, HD~188041) the formal solution for the Ca distribution yields an unrealistically high Ca
abundance in the lower atmosphere. However, we find that in these objects the actual inferred
stratification profile often has a plateau with $\log (Ca/N_{\rm tot})\approx-3.5$ in the layers
around $\log\tau_{5000}\approx0$.
This value of the abundance gives the actual element concentration that influences
the Ca line formation. The formal high Ca abundance occurs far below the photosphere (e.g., the
bottom panel of Fig.~\ref{Ca_distr} shows this situation for HD~137949). It does not influence 
the observed Ca line profiles. Therefore, for all stars marked by a letter '$c$' in
Table~\ref{ca_strat_tbl} we took the Ca abundance at $\log\tau_{5000}\approx0$ as 
$\log (Ca/N_{\rm tot})_{\rm lo}$, modifying the position and the width of the abundance jump
correspondingly.} 

Homogeneous determination of the calcium vertical distribution in a large number of magnetic
Ap stars allows us to search for possible dependence of diffusion signatures on the stellar
atmospheric parameters and magnetic field strength. However, except of the marginal tendency of
the Ca abundance in the upper layers to be higher in hotter stars, we found no clear
correlations between each of the four stratification parameters on the one hand and \teff\ and
\bs\ on the other hand. Although the qualitative form of the Ca stratification is the same in
all stars, possibly indicating the action of a universal radiative diffusion process, the
observed star-to-star variation in stratification profiles appears to be irregular and unrelated
to any other stellar characteristic.

\subsection{The Ca isotopic anomaly}

\begin{figure*}[!t]
\centering
\figgps{82mm}{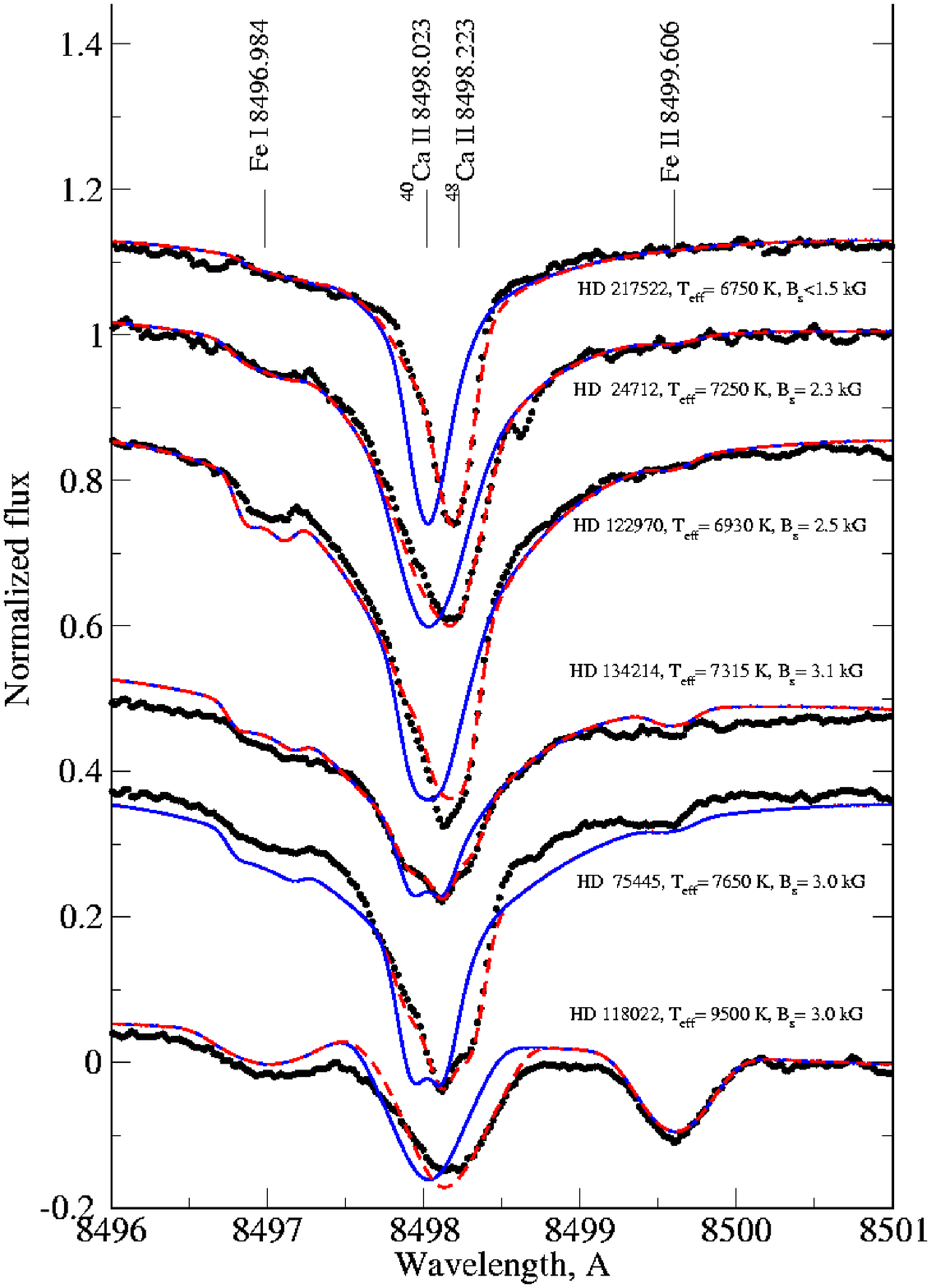}\hspace{0.5cm}\figgps{82mm}{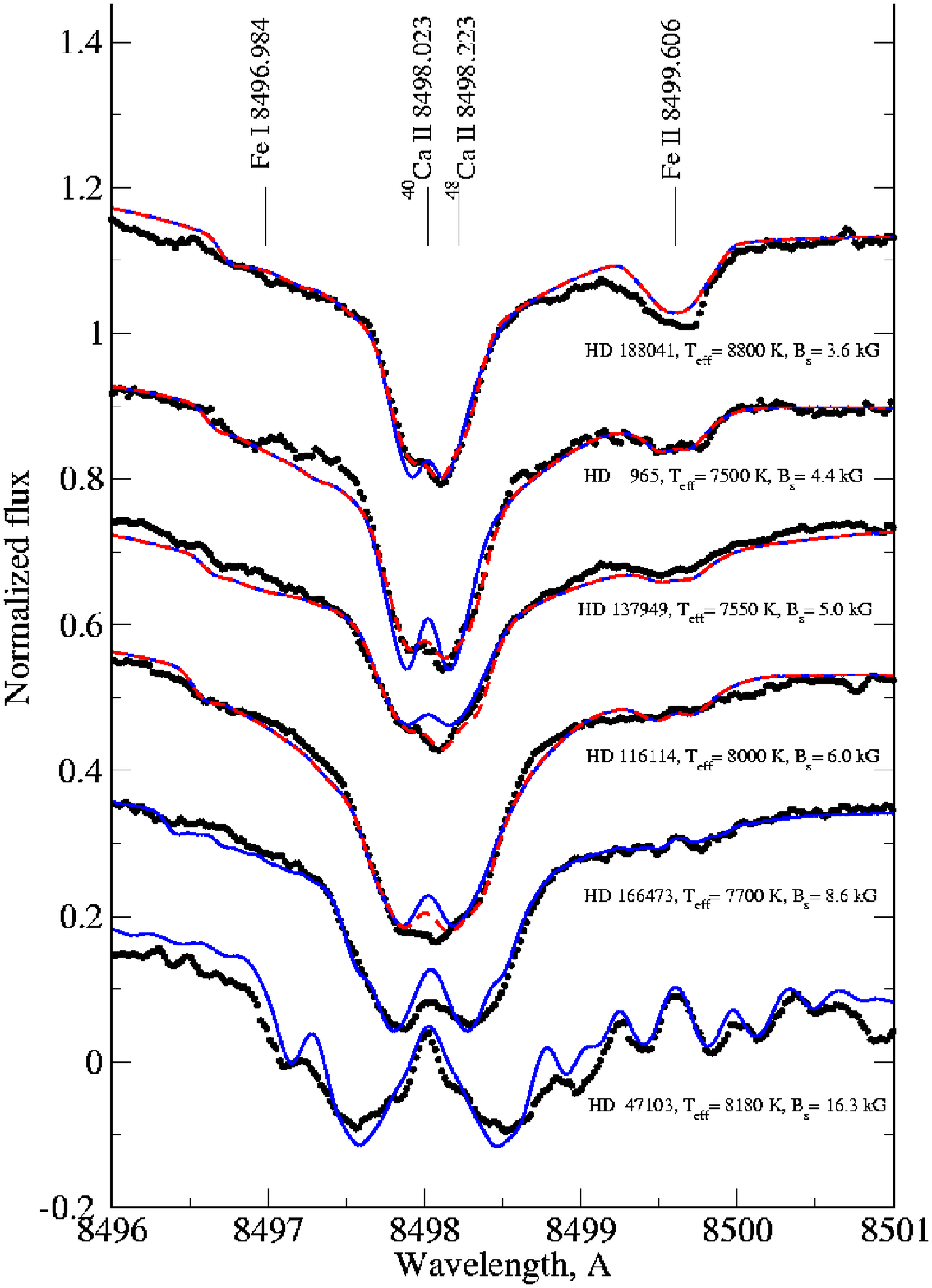}
\caption{Comparison of the observed (full circles) and calculated (lines) profiles of the
\ion{Ca}{ii} 8498\,\AA\ line in a subset of program stars. Solid line shows theoretical
calculations for the terrestrial Ca isotopic mixture and the Ca stratification derived
from optical lines. The dashed line presents spectrum synthesis for the models
with vertical separation of the Ca isotopes. The stars are arranged in order of increasing 
magnetic field strength.}  
\label{8498_all}
\end{figure*}
 
The Ca isotopic analysis procedure outlined in Sect.~\ref{isot} was applied to all stars
included in our paper. Table~\ref{ca_strat_tbl} lists the lower boundary 
\txtbf{(uncertain to within $\approx$\,0.1~dex)} of the heavy Ca
layer in the atmospheres of magnetic Ap stars for which the \ion{Ca}{ii} 8498\,\AA\ line 
could not be reproduced with the terrestrial isotopic mixture. Our estimate of the 
dominant contributor among
different heavy Ca isotopes is also given. Out of 23 program stars, the
presence of heavy isotopes is established in 17 objects.

For three stars, HD\,965, HD\,137909, HD\,137949, the modelling of the IR
triplet line requires an increase of the Ca abundance in the upper atmosphere, above the
abundance jump given by the simple stratification analysis. In these cases we attribute the
high-lying cloud to heavy Ca. 

Figure~\ref{8498_all} compares theoretical spectrum synthesis calculations with observations
for a subset of 12 cooler (\teff\,$\le$\,9000~K) stars with different effective temperatures 
and different magnetic field intensity.
The stars are arranged in order of increasing magnetic field strength. It is evident that
the presence of heavy Ca isotopes inversely correlates with magnetic field strength. 
This effect is further illustrated in Fig.~\ref{isall}a, where we quantify
the detection of heavy Ca by the ratio of calculated equivalent widths of the 
$^{44}$Ca+$^{46}$Ca+$^{48}$Ca to $^{40}$Ca line components. In the stars
with small to moderate magnetic fields we clearly see a significant contribution of the
heavy isotopes $^{46}$Ca and $^{48}$Ca (large equivalent width ratio). This contribution decreases with the increase of
magnetic field strength. Even in HD~137909 ($\beta$~CrB) with the mean magnetic field modulus
\bs\,=\,5.4~kG one still needs a small contribution of $^{48}$Ca, but under the assumption of
very specific Ca distribution shown in Fig.~\ref{Ca_distr}. 
No evidence for the presence of heavy Ca is found in cool Ap stars with \bs\,$\ga$6~kG. 

\begin{figure}[!th]
\centering
\figgps{82mm}{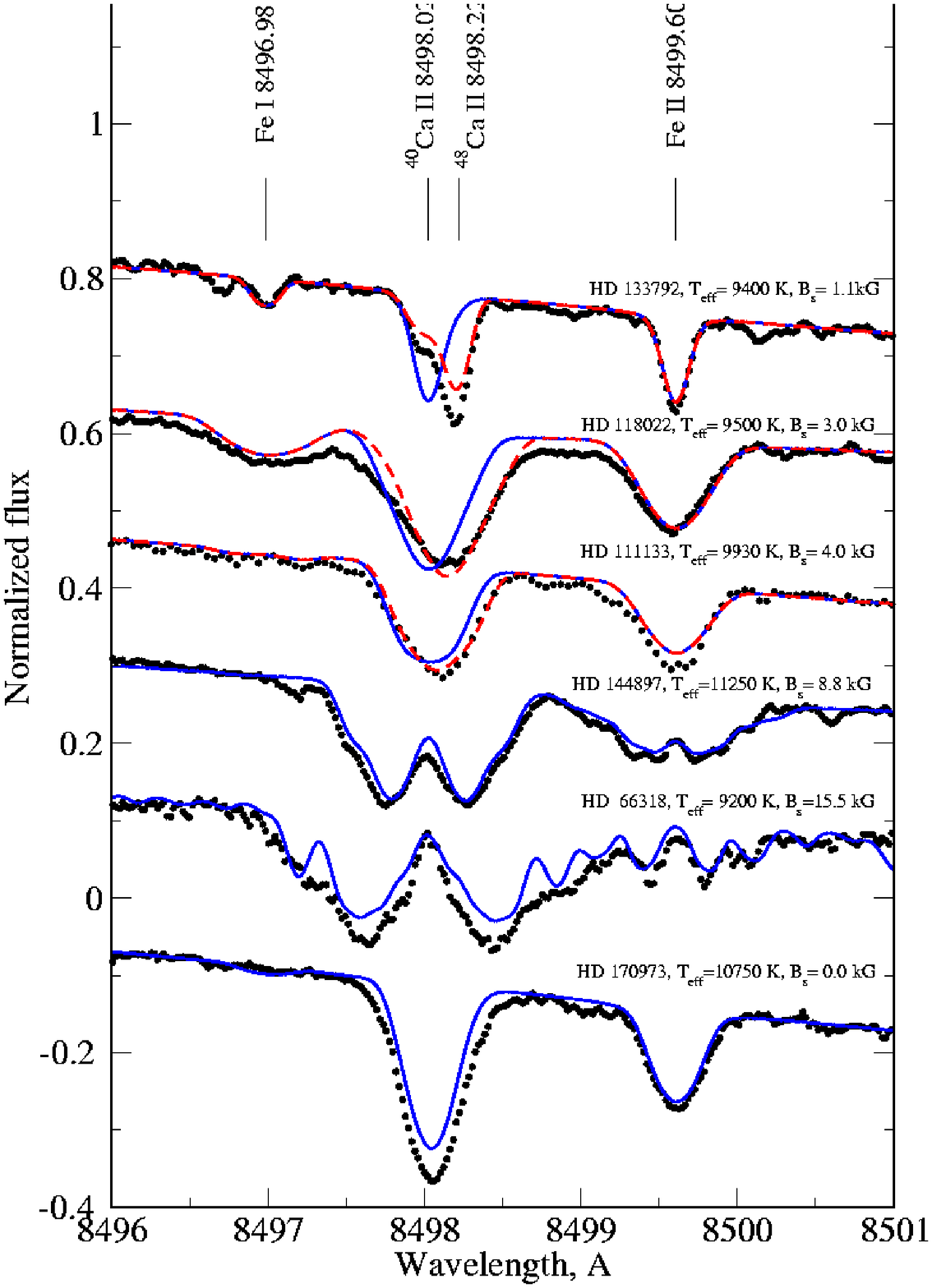}
\caption{Same as in Fig.~\ref{8498_all} but for the stars with \teff$>$ 9000 K.}  
\label{8498_hot}
\end{figure}

\begin{figure*}[!th]
\centering
\figgps{15cm}{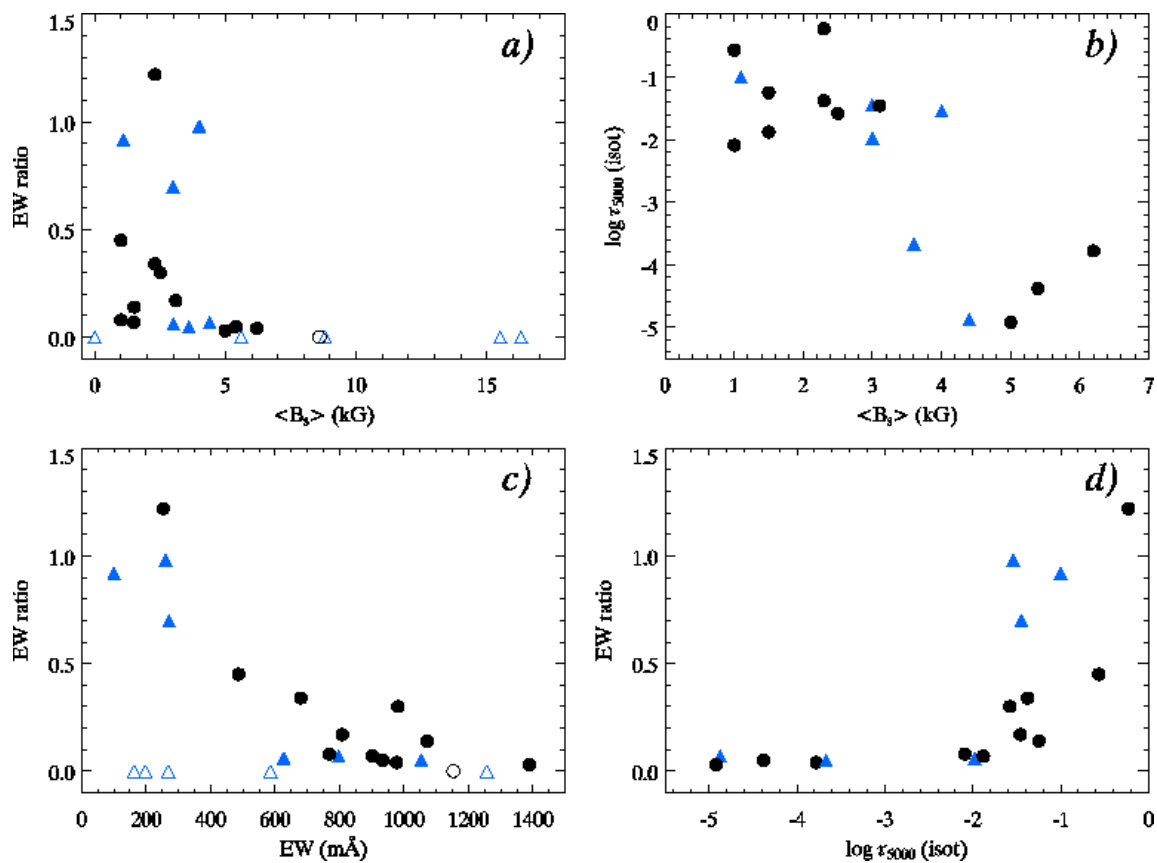}
\caption{Results of the analysis of Ca isotopic composition in Ap stars. 
{\bf a)} ratio of calculated equivalent width of the heavy and light Ca
components of the 8498\,\AA\ line vs. magnetic field strength. 
{\bf b)} Position of the boundary between the light and heavy Ca layers
vs. magnetic field strength. 
{\bf c)} The equivalent width ratio vs. the total equivalent width of
the \ion{Ca}{ii} 8498\,\AA\ line. 
{\bf d)} The equivalent width ratio vs. the optical depth of the boundary
between the light and heavy Ca layers.
In all plots different symbols correspond to Ap stars with detected presence of heavy Ca
({\it filled symbols}) and those showing no heavy Ca isotopes ({\it open symbols}), as well
as to roAp ({\it circles}) and non-pulsating Ap ({\it triangles}) stars.}  
\label{isall}
\end{figure*}

Due to the temperature dependence of the Ca line intensity, we used a smaller number of  lines for
stratification analysis of hot Ap stars (\teff\,$>$\,9000~K). In these stars  stratification is
defined mainly by the \ion{Ca}{ii} lines. Among the neutral Ca lines only the  resonance
\ion{Ca}{i}~4227\,\AA\ line could be utilized. The agreement of calculated and observed line
profiles is poorer for hot stars. However, the general anticorrelation between the presence of the
heavy Ca isotopes and magnetic field strength is  still apparent (see Fig.~\ref{8498_hot}). As in
cool stars, no evidence for the presence of heavy Ca is found in hot Ap stars with \bs\,$\ga$6~kG. 
The only exception is HD~170973, which has no observable magnetic field and shows no contribution of
the heavy Ca isotopes. We note that HD~170973 is also the only star for
which we find no evidence of Ca stratification. 
This star also possesses the highest Ca abundance among the hot Ap stars in our sample.

Thus, it appears that the mechanism
responsible for the accumulation of heavy Ca isotopes in the line-forming atmospheric region
works more efficiently in stars with a small magnetic field strength. 
Figure~\ref{isall} illustrates several other interesting relations between the Ca isotopic
anomaly and stellar parameters that we have revealed in this study. The optical depth of
the layer separating the heavy and light Ca clouds depends on the magnetic field strength.
In stars with more intense fields heavy
Ca isotopes tend to accumulate much higher in the stellar atmospheres
(Fig.~\ref{isall}b). There is also an anticorrelation of the heavy Ca contribution and the total
equivalent width of the \ion{Ca}{ii} 8498\,\AA\ transition (Fig.~\ref{isall}c), which implies
that the efficiency of the isotope separation process depends on the total line intensity in
cases where such separation exists.

We find no difference between roAp and non-pulsating Ap stars (shown with different
symbols in Fig.~\ref{isall}) regarding the presence of heavy Ca excess in their atmospheres
and its correlation with the magnetic field strength and line intensity.

\section{Discussion and conclusions}
\label{discus}

In this paper we have investigated the vertical stratification and isotopic
anomaly of Ca in a large sample of cool magnetic Ap stars. Our study is the first to address
the interesting problem of the presence of heavy Ca isotopes in chemically peculiar
stars with detailed polarized radiative transfer calculations, which take into account the
effects of the magnetic field and chemical
separation in stellar atmospheres. We derive stratification of Ca for 23 Ap
stars using a sample of \ion{Ca}{i} and \ion{Ca}{ii} lines distributed over a broad spectral
range. All but one program stars clearly show signatures of the Ca stratification, whereas 
comparison stars reveal no inhomogeneities in the vertical Ca distribution when analysed with the
same techniques and atomic data. 
Although our stratification modelling was based on a simplified step-function approximation
and adopted a simplified homogeneous model for the magnetic field geometry, 
the inferred parameters of the vertical Ca distributions impose
important observational constraints for theoretical models of the radiative diffusion processes in
stellar atmospheres.

Analysis of the IR Ca triplet line \ion{Ca}{ii} 8498\,\AA\ provided information on the
relative contribution of different Ca isotopes. Spectrum synthesis calculations show that
significant isotopic shifts observed in the core of \ion{Ca}{ii} 8498\,\AA\ cannot be
attributed to the overabundance of heavy Ca isotopes throughout the whole stellar
atmosphere.
Instead, \txtbf{we show that} the Ca line profile shape \txtbf{is consistent with} the vertical
separation of different Ca isotopes, with heavy Ca located in a cloud above the most
abundant isotope $^{40}$Ca. Even though the presence of heavy Ca is prominent in the line
core, the normal Ca isotope dominates the line wings and is more abundant in the lower
atmosphere where the total Ca abundance is also much larger. Thus, our \txtbf{tentative} model calls for a
less extreme heavy Ca enrichment than was suspected in previous investigations limited to
the centroid measurements of the \ion{Ca}{ii} 8498\,\AA\ line core.

Recently different stratification of He isotopes was found by Bohlender (\cite{B05}) 
in the analysis of $^{3}$He and related stars. That paper used a similar approach to
modelling chemical stratification, but has only addressed He vertical distribution in hot 
non-magnetic chemically peculiar stars.

We have successfully used the model with Ca stratification and vertical isotopic separation
to explain the appearance of the \ion{Ca}{ii} 8498\,\AA\ in all stars showing excess of
heavy Ca isotopes. The prominent anticorrelation between the presence of heavy Ca and magnetic
field strength, first reported by Ryabchikova (\cite{MONS05}), is confirmed and strengthened
in the present paper. We find that only stars with sufficiently weak field show traces of
heavy Ca. According to our knowledge, this interesting relation is the only case when
definite dependence of the chemical abundance characteristic on the magnetic field strength
is found for Ap stars. Furthermore, we find that in stars with stronger fields the heavy Ca
isotopes tend to accumulate higher in the atmosphere. 

According to the results of our study, pulsating (roAp) and non-pulsating Ap stars are not
distinguished by the characteristics of their Ca isotopic anomaly. Both groups of stars are
equally likely to show an excess of heavy Ca and follow the same trend of the heavy isotope
contribution versus the magnetic field strength and the \ion{Ca}{ii} 8498\,\AA\ line intensity.

If the overall distribution of Ca abundance in the atmospheres of Ap stars follows the
predictions of the radiatively driven diffusion, our results on the isotopic separation
favour the light-induced drift (LID) as the main process responsible for  this separation.
According to Atutov \& Shalagin (\cite{AS88}), LID arises when the radiation field is
anisotropic inside the line profile. Such an anisotropy takes place for a line of the trace
isotope, for instance $^{46}$Ca and $^{48}$Ca in the terrestrial calcium mixture, which is
located in the wing of a strong line of the main isotope $^{40}$Ca. The main isotope induces
the drift velocity for other isotopes. If the trace isotope's line is located in the red
wing of the line due to main isotope, the drift velocity is directed towards the upper
atmosphere and the trace isotopes are pushed upwards. This is in agreement with observed
vertical distribution of Ca isotopes. The Zeeman splitting changes the line shape and
decreases the flux anisotropy for the line of trace isotope. When magnetic field becomes
strong enough, $\sim$5--6~kG, the flux anisotropy disappears and the isotopic separation is
ceasing. Therefore, the observed Ca isotopic anomaly in magnetic stars may be qualitatively
explained by the combined action of the radiatively-driven diffusion and the light-induced
drift. 
Theoretic study of a combination of the radiatively-driven diffusion and LID was
presented by LeBlanc \& Michaud (\cite{LM93}) for He. These authors showed
that LID accelerates separation of $^{3}$He from $^{4}$He in hotter CP
stars.
Detailed theoretical chemical diffusion calculations
(LeBlanc \& Monin \cite{LM04}) should incorporate LID in order to test our hypothesis 
that this effect may be important for the chemical transport processes in cool Ap-star 
atmospheres.

\begin{acknowledgements}
We thank Luca Fossati for letting us use his spectrum of HD\,73666. 
This work was supported by the RAS Presidium Program ``Origin and Evolution of Stars and
Galaxies'', by Austrian Science  Fund (FWF-P17580N2) and by the grant 11630102 from the Royal
Swedish Academy of Sciences. TR also acknowledges partial support from the RFBR grant
06-02-16110a and the Leading Scientific School grant 162.2003.02.
OK acknowledges financial support from the Swedish Royal Physiographic Society 
in Lund.
\end{acknowledgements}

\end{document}